\documentclass[aps,prb,twocolumn,showpacs,preprintnumbers,amsmath,amssymb]{revtex4}

\usepackage{graphicx}
\usepackage{dcolumn}
\usepackage{bm}
\usepackage[latin1]{inputenc}
\usepackage{psfrag}
\usepackage{color}
\usepackage{hhline}

\begin{document}

\title{Surface induced magnetization reversal of MnP nanoclusters embedded in GaP}

\author{Christian Lacroix}
\email{christian.lacroix@polymtl.ca}
\author{Samuel Lambert-Milot}
\author{Patrick Desjardins}
\author{Remo A. Masut}
\author{David Ménard}
\email{david.menard@polymtl.ca}

\affiliation{%
Department of Engineering Physics and Regroupement québécois sur les matériaux de pointe (RQMP), Polytechnique Montréal, H3T 1J4, Canada
}%

\pagebreak

\begin{abstract}
\noindent We investigate the quasi-static magnetic behavior of ensembles of non-interacting ferromagnetic nanoparticles consisting of MnP nanoclusters embedded in GaP(001) epilayers grown at 600, 650 and 700°C. We use a phenomenological model, in which surface effects are included, to reproduce the experimental hysteresis curves measured as a function of temperature (120-260 K) and direction of the applied field. The slope of the hysteresis curve during magnetization reversal is determined by the MnP nanoclusters size distribution, which is a function of the growth temperature. Our results show that the coercive field is very sensitive to the strength of the surface anisotropy, which reduces the energy barrier between the two states of opposite magnetization. Notably, this reduction in the energy barrier increases by a factor of 3 as the sample temperature is lowered from 260 to 120 K.
\end{abstract}
          
                        
\maketitle

\section{Introduction}
With recent advances in nanotechnologies, magnetic nanoparticles (MNPs) or nanomagnets embedded in a nonmagnetic matrix may soon become part of novel applications, such as ultra-high-density recording media~\cite{sun} or optoelectronic technologies.~\cite{mayergoyz2012} Therefore, it becomes of practical as well as fundamental interest to understand the mechanisms behind the magnetization reversal of these MNPs. 

The role of the surface (or the interface) on the magnetization reversal of MNPs becomes important as the size decreases and has been recognized and investigated both experimentally~\cite{chen1998,jamet2001,tournus2010} and theoretically~\cite{zhang1996,nakatani2000,yanes2009}. In terms of magnetic stability, the surface anisotropy can be used to increase the energy barrier between the two states of opposite magnetization. As a consequence, it reduces the probability of magnetization switching due to thermal fluctuations as the size of the magnetic domains diminished.~\cite{chen1998}

In diluted granular systems, the dipolar and exchange interactions between the nanomagnets are negligible. Generally, the zero field cooled-field cooled protocol along with the magnetization curves above the blocking temperature are used to extract the surface anisotropy contribution and size distribution.~\cite{binns2005,Tamion2009} However, no analytical modeling of the complete hysteresis curve of diluted granular systems below the blocking temperature, where the coercive field is not zero, has ever been reported.

Here, we propose modifications to the well known Stoner-Wohlfarth model to take into account a orthorhombic magnetic anisotropy, and incorporate a surface (interface) contribution that modifies the energy barrier between the two states of opposite magnetization. Our model allow to reproduce the experimental hysteresis curve of MnP nanoclusters embedded in a GaP epilayer. X-ray diffraction (XRD) and ferromagnetic resonance (FMR) measurements have shown that the MnP clusters are orthorhombic with their \textit{c}-axis (3.173 \AA) oriented along the GaP $\left\langle 110\right\rangle$ directions, \textit{b}-axis (5.260 \AA) along GaP $\left\langle 111\right\rangle$ or $\left\langle 001\right\rangle$ and \textit{a}-axis (5.917 \AA) along GaP $\left\langle 11\bar{2}\right\rangle$ or $\left\langle 110\right\rangle$.~\cite{lacroix2013,lambert2012} Careful analysis of XRD and FMR data allowed us to determine the volume fraction of MnP corresponding to each GaP crystallographic orientation. Cross-sectional and plane-view TEM micrographs reveal that the MnP nanoclusters are of quasi-cylindrical shape whose major to minor axis ratio is approximately 1.3.~\cite{lambert} The MnP nanoclusters are distributed uniformly throughout the epilayer and occupy between 4 and 7\% of the epilayer's volume depending up on growth conditions.

\section{Experimental procedure}

A series of gallium phosphide epilayers with embedded MnP nanoclusters (GaP:MnP) whose thicknesses vary between 1 and 1.3 $\mu$m were grown on semi-insulating GaP(001) substrates using a low-pressure, cold-wall, MOVPE reactor.~\cite{lambert} Samples were deposited on GaP buffer layers using three different substrate temperatures: 600, 650 and 700°C. We call these GMP600, GMP650 and GMP700 respectively. 

To determine the nanocluster size distribution of a given sample, the dimensions of 100 MnP nanoclusters were measured from a cross-sectional TEM micrograph. We assumed that the depth of the nanoclusters (dimension normal to the TEM micrograph) was equal to the smallest dimension of the nanocluster as measured in the micrograph plane. 

\begin{figure} 
\begin{center}
\includegraphics[width=0.5\textwidth]{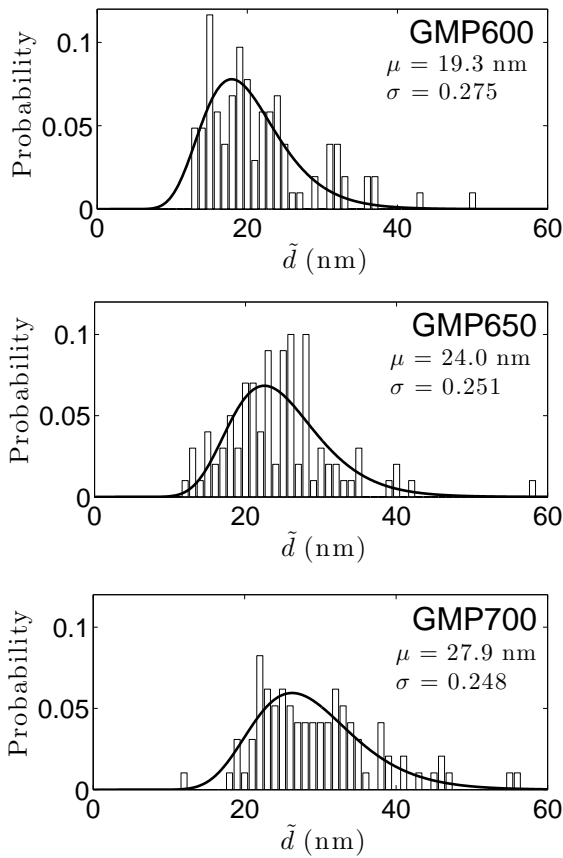}
\caption{\label{fig:TEM} Size distribution obtained from TEM images for samples GMP600, GMP650 and GMP700. Solid curves correspond to a log-normal distribution fit whose parameters are indicated in the plots.}
\end{center}
\end{figure}

Hysteresis curves of the GaP:MnP epilayers were measured along different crystallographic directions relative to the GaP substrate using a vibrating sample magnetometer, at temperatures ranging from 120 to 260 K. The samples were mounted on a quartz rod. The magnetic moment of the GaP:MnP epilayers was obtained after subtracting the magnetic moment of the substrate. The resulting value was then divided by the volume of the epilayer. 

\section{Results}

Processing of the TEM images yields the histograms presented in Fig.~\ref{fig:TEM} where the probability $g_\textrm{TEM}(\tilde{d})$ is plotted as a function of $\tilde{d}$, where $\tilde{d} = (6V/\pi)^\frac{1}{3}$. The distributions can be fitted using a log-normal curve, \textit{i.e.}

\begin{eqnarray}
g_\textrm{TEM}(\tilde{d}) = \frac{1}{\sqrt{2\pi}\sigma\tilde{d}}e^\frac{{-\left(\ln\left(\tilde{d}/\mu\right)\right)^2}}{2\sigma^2},
\label{eq:lognormal}
\end{eqnarray}

\noindent where $\mu$ and $\sigma$ are the mean of $\tilde{d}$ and standard deviation of $\ln\tilde{d}$ respectively. We observe that, as the growth temperature increases, the mean diameter increases.

\begin{figure} 
\begin{center}
\includegraphics[width=0.5\textwidth]{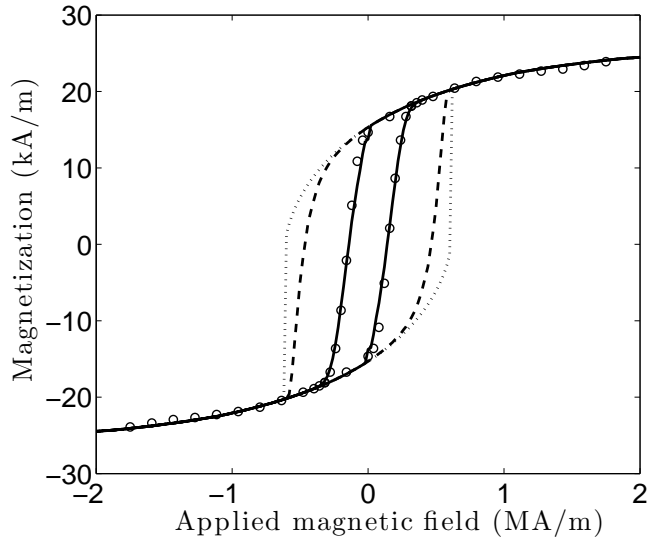}
\caption{\label{fig:hystnoHs} Experimental hysteresis curve of sample GMP600 (circles) measured at $T$ = 180 K when the magnetic field is applied in the GaP $[001]$ direction (out of sample plane). The solid line was obtained using the model described in the text and includes the surface anisotropy as well as thermal fluctuations. The only fitting parameters used in the calculations are the surface anisotropy constant $K_s$ and the volume fraction $V_n/V_e$ occupied by the MnP nanoclusters (see Table~\ref{tab:param}). For comparison, the dashed line was obtained for the case where $K_s = 0$ but thermal fluctuations are present. The dotted line corresponds to the hysteresis curve obtained assuming $\Delta E_\textrm{surf} = \Delta E_t = 0$.}
\end{center}
\end{figure}

\begin{figure*} 
\begin{center}
\includegraphics[width=1\textwidth]{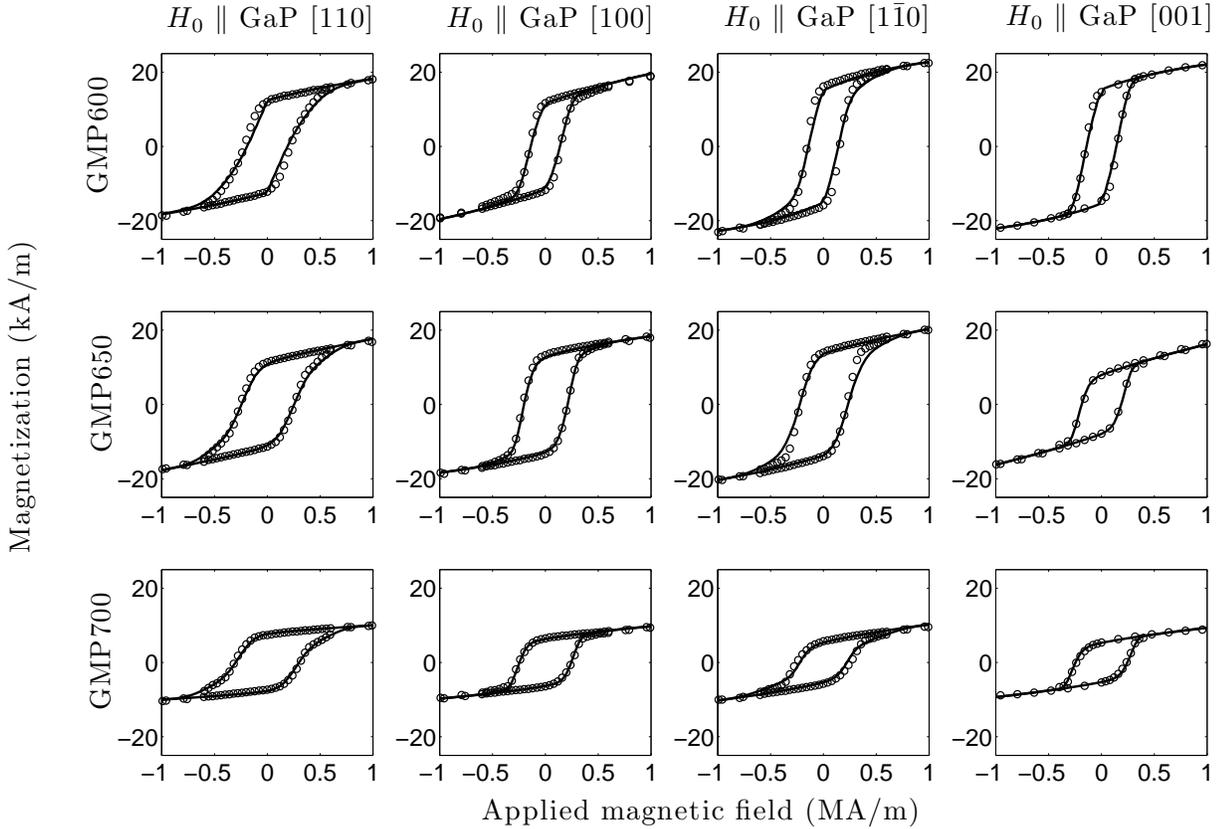}
\caption{\label{fig:hystGMP} Experimental hysteresis curves (circles) measured at $T$ = 180 K for samples GMP600, GMP650 and GMP700  when the magnetic field is applied in the GaP $[110]$ direction (in-plane), GaP $[100]$ (in-plane), GaP $[1\bar{1}0]$ (in-plane) and GaP $[001]$ (out of plane). Each column corresponds to a specific field direction relative to the GaP substrate while each row corresponds to a specific sample. Solid lines were calculated using the model described in the text, which includes the effect of the surface anisotropy and thermal fluctuations on the energy barrier. The values of $V_n/V_e$ and $K_s$ used in the calculations can be found in Table~\ref{tab:param}. }
\end{center}
\end{figure*}

The hysteresis curves of the samples, measured at $T$~=~180 K with the magnetic field applied in various GaP crystallographic directions, are presented in Figs.~\ref{fig:hystnoHs}~and~\ref{fig:hystGMP} (circles). The measurement time for each data point was $t_\textrm{meas}$ = 100 s. We observe that the hysteresis curve depends not only up on the sample but also up on the direction of the applied magnetic field.

\section{Model}

\begin{figure} 
\begin{center}
\includegraphics[width=0.4\textwidth]{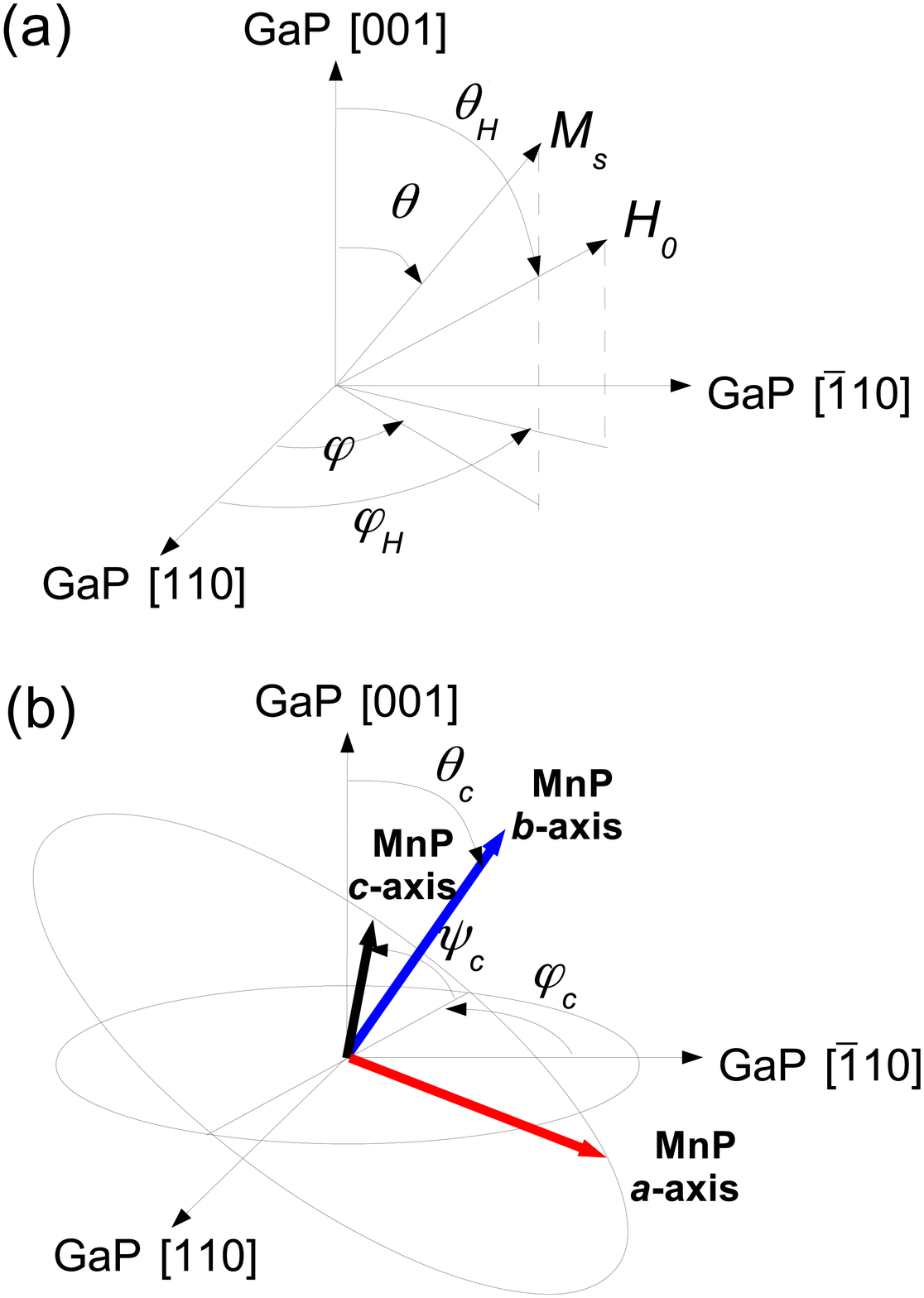}
\caption{\label{fig:coordsystem} (a) The direction of the static field $H_0$ and the magnetization $M_s$ relative to the GaP substrate are described by the angles ($\theta_H$, $\varphi_H$) and ($\theta$, $\varphi$) respectively. (b) The orientation of the MnP magnetocrystalline anisotropy relative to the GaP substrate is described by the angles $\psi_c$, $\theta_c$ and $\varphi_c$. }
\end{center}
\end{figure}

We treat the MnP nanoclusters as ferromagnetic single domains (macrospins) which have a saturation magnetization $M_s$ (equal to the spontaneous magnetization). This means that, for a constant temperature, the magnetization $\vec{M}$ of a cluster will be considered to be a vector of constant length $M_s$. The magnetization direction will be defined relative to the crystallographic orientation of the GaP substrate using the angles $\theta$ and $\varphi$, while the angles $\theta_H$ and $\varphi_H$ define the direction of the applied magnetic field $H_0$ relative to the crystallographic orientation of the GaP substrate [see Fig.~\ref{fig:coordsystem}(a)].

In order to describe the magnetic state of a MnP nanocluster, it is appropriate to use the magnetic contribution to its free energy density. The magnetic contribution $U$ is the sum of the Zeeman energy density $U_Z$ and a orthorhombic magnetic anisotropy term $U_a$ dominated by the magnetocrystalline anisotropy of MnP. The demagnetizing energy due to clusters' shape can be neglected as demonstrated in Ref.~\onlinecite{lacroix2013}.

The Zeeman energy density $U_Z$ is expressed as: 

\begin{widetext}
\begin{eqnarray}
	U_Z &=&  -\mu_0M_sH_0\left[\sin\theta \sin\theta_H \cos\left(\varphi-\varphi_H \right)+ \cos\theta \cos\theta_H\right]
	\label{eq:UZ},
\end{eqnarray}

\noindent and the energy density for the case of a orthorhombic anisotropy is expressed as, 

\begin{eqnarray}
U_a = K_1A^2 + K_2B^2,
\label{eq:Ua}
\end{eqnarray}

\noindent where 

\begin{eqnarray}
A &=& \sin\theta\cos\psi_c\cos\theta_c\cos\left(\varphi-\varphi_c\right) + \sin\theta\sin\psi_c\sin\left(\varphi-\varphi_c\right)-\cos\theta\cos\psi_c\sin\theta_c, \nonumber \\
B &=& \sin\theta\sin\theta_c\cos\left(\varphi - \varphi_c\right) + \cos\theta\cos\theta_c,
 \nonumber
\end{eqnarray}
\end{widetext}

\noindent $\psi_c$, $\theta_c$ and $\varphi_c$ are the angles used to describe the orientation of the magnetic anisotropy, which corresponds to the magnetocrystalline anisotropy of bulk MnP, relative to the crystallographic orientation of the GaP substrate [see Fig.~\ref{fig:coordsystem}(b)]. 

In order to calculate the dependence of the magnetization with the applied magnetic field, the first and second derivatives of $U$ [Eqs.~(\ref{eq:UZ})-(\ref{eq:Ua})] are used to find the minimum of $U$ and thus, the direction of the magnetization. Except when the magnetic field is applied perpendicular to the easy axis, two relative minima are found when the applied field magnitude is below a threshold value, the critical field, corresponding to two directions of the magnetization. The projections of these two magnetization directions on the direction of the applied magnetic field give the hysteresis curve and will be called $m_+$ and $m_-$. These two directions correspond to two stable magnetization states, where an energy barrier $\Delta E$ must be overcome so that the magnetization can switch from one stable state to the other state. As a consequence, the magnetic response shows hysteretic behaviour.

It is well known that thermal fluctuations reduce the critical field and that very fine ferromagnetic particles can even become superparamagnetic,~\cite{neel,brown} \textit{i.e.} that their magnetization can change direction spontaneously, after a time $\tau_\textrm{sw}$, which we call the switching time. Thus, if we measure the magnetization of a very fine ferromagnetic particle over a measurement time $t_\textrm{meas} \gg \tau_\textrm{sw}$, the measured value will be zero and the particle is considered superparamagnetic. However, if $t_\textrm{meas} \ll \tau_\textrm{sw}$, the measured value will be non zero and the particle is considered ferromagnetic. 

We assume that the switching time $\tau_\textrm{sw}$ can be calculated using the Arrhenius expression,~\cite{bean}

\begin{eqnarray}
\tau_\textrm{sw} = \tau_0e^{\frac{\Delta E}{k_BT}},
\label{eq:superpara}
\end{eqnarray} 

\noindent where $\tau_0$ is the time between two attempts of the particle's magnetization to change direction, $\Delta E$ is the energy barrier and depends up on the magnetic field and the particle's size, $k_B$ is the Boltzmann constant and $T$ is the particle's temperature. The pre-exponential factor $\tau_0$ can be estimated using $\tau_0^{-1} \approx \left(2\pi\right)^{-1}\gamma\mu_0 H_\textrm{loc}$, where $\gamma$ is the gyromagnetic ratio and $H_\textrm{loc}$ is the local magnetic field \textit{seen} by the magnetic atoms, which is expressed as

\begin{equation}
H_\textrm{loc} =\frac{1}{\mu_0M_s \sin \theta_0} \left[\frac{\partial^2U}{\partial\theta^2}\frac{\partial^2U}{\partial\varphi^2} - \left(\frac{\partial^2U}{\partial\theta\partial\varphi}\right)^2\right]^{1/2}_{\theta_0,\varphi_0}.
\label{eq:smitsuhl}
\end{equation}

If we replace $\tau_\textrm{sw}$ by $t_\textrm{meas}$ in Eq.~(\ref{eq:superpara}), we then find the height of the energy barrier that thermal fluctuations can overcome. We can also interpret this as follows: the effect of thermal fluctuations is to reduce the energy barrier $\Delta E$ between the two states of opposite magnetization by the amount $\Delta E_{t} = k_BT\ln{(t_\textrm{meas}/\tau_0)}$.

It has been demonstrated earlier, through micromagnetic simulations, that in the presence of a surface anisotropy and in the case where the core spins are coupled with the surface spins through an exchange interaction, the critical field, and therefore the energy barrier, can be reduced or increased.~\cite{kachkachi2002} The increase or reduction depends on the relative strengths of the surface anisotropy, volume anisotropy and exchange interaction between the surface spins and core spins. Here, we postulate the presence of a surface-induced reversal mechanism that lowers the energy barrier. The energy barrier thus becomes

\begin{eqnarray}
	\Delta E (\tilde{d},H_0) = \Delta E_\textrm{SW}(\tilde{d},H_0) - \Delta E_\textrm{surf}(\tilde{d}) - \Delta E_t
	\label{eq:deltaE}
\end{eqnarray}

\noindent where $\Delta E_\textrm{SW} = K_\textrm{eff}V$ is the energy barrier as calculated from the Stoner-Wohlfarth model, $K_\textrm{eff}$ is an effective anisotropy constant and $\Delta E_\textrm{surf}$ is a term related to the surface anisotropy of the MnP nanoclusters. We note that $\Delta E_\textrm{surf} = K_s\pi\tilde{d}^{2}$, where $K_s$ is a surface anisotropy constant. In our model, for a given sample and temperature, we use a single surface anisotropy constant $K_s$ for all nanoclusters. As discussed in Ref.~\onlinecite{tournus2010}, the dominating factor in the energy barrier distribution is often the size distribution, rather than a distribution of $K_s$ values. In the case where $\Delta E (\tilde{d},H_0 = 0) < 0$, the particle's magnetization can switch spontaneously between the two stable magnetization states. The particle's contribution to the total magnetization of the epilayer is then zero, \textit{i.e.} the particle is superparamagnetic.

To include the effects of thermal fluctuations and surface anisotropy in the hysteresis curve, we first calculate the energy barrier $\Delta E_\textrm{SW}$ using Eqs.~(\ref{eq:UZ})-(\ref{eq:Ua}). As $H_0$ increases, $\Delta E_\textrm{SW}$ becomes smaller or larger depending up on the field direction relative to the direction of the magnetization. In the case where the direction of $\vec{H_0}$ projected on $\vec{M}$ is opposite to the direction of $\vec{M}$, $\Delta E_\textrm{SW}$ diminishes as the applied field increases and we eventually reach the critical field, \textit{i.e.} $\Delta E_\textrm{SW}$ becomes zero, meaning that only one stable state (instead of two) is present. As a consequence, the magnetization switches into the only stable state. If we replace $\Delta E_\textrm{SW}$ by $\Delta E$ [Eq.~(\ref{eq:deltaE})], the critical field is then reduced. This implies a reduction of the coercive field in the hysteresis curve. 

\begin{widetext}

To fully model the hysteresis curve of the GaP:MnP epilayers, we used several approximations. Considering that the volume fraction occupied by the nanoclusters in the epilayer is relatively low ($\approx$ 4 - 7\%) and that MnP possesses a strong magnetocrystalline anisotropy, we can neglect the dipolar interaction between the clusters.~\cite{lacroix2013} Therefore, we treat the MnP nanoclusters embedded in a GaP matrix as an ensemble of non-interacting ferromagnetic single domains having a log-normal type distribution size $g_\textrm{TEM}$ and whose crystallographic orientation is described by a distribution function $f$. Assuming also that all the MnP nanoclusters have the same magnetization $M_s$, the total magnetization of the GaP:MnP epilayer can be expressed as

\begin{eqnarray}
M\left(H_0, \theta_H, \varphi_H, T \right) = M_s\left(T\right)\frac{V_n}{V_e}\int_{\psi_c}\int_{\theta_c}\int_{\varphi_c} \int_{\tilde{d}}m\left(H_0,\theta_H, \varphi_H, \varphi_c, \theta_c, \psi_c, \tilde{d}\right) g_\textrm{TEM}(\tilde{d}) f\left(\varphi_c, \theta_c, \psi_c\right)d\tilde{d}d\varphi_c d\theta_c d\psi_c
\label{eq:Mtot}
\end{eqnarray}

\noindent where $V_n/V_e$ is the volume fraction occupied by the nanoclusters in the epilayer, $V_n$ is the volume occupied by the nanoparticles, $V_e$ is the total volume of the epilayer, $g_\textrm{TEM}$ is the size distribution obtained from TEM images and $m = m_+$ or $m_-$, which are the projections of the two stable magnetization directions on the direction of the applied magnetic field as described earlier. In the case where $\Delta E (\tilde{d},H_0 = 0)$ is negative, $m = 0$.

Knowing that the nanoclusters are oriented in a finite number of preferred crystallographic directions relative to the GaP substrate and neglecting the angular dispersion $\Delta\varphi_c$, $\Delta\theta_c$ and $\Delta\psi_c$ around each direction, then

\begin{eqnarray}
f\left(\varphi_c, \theta_c, \psi_c\right) = \sum^n_{i=1} a_i\delta\left(\varphi_c-\varphi_{ci}\right)\delta\left(\theta_c-\theta_{ci}\right)\delta\left(\psi_c-\psi_{ci}\right)
\label{eq:f},
\end{eqnarray}

\noindent where $a_i = V_{i}/V_n$ is the volume fraction of MnP nanoclusters oriented along the direction $i$, $V_{i}$ is the total volume of MnP nanoclusters oriented along the direction $i$ and $n$ is the number of orientations.~\footnote{In the case of our GaP:MnP samples, $n$ = 18 (see Ref.~\onlinecite{lacroix2013}).} Equation~(\ref{eq:Mtot}) can be expressed as

\begin{eqnarray}
M\left(H_0, \theta_H, \varphi_H, T \right) = M_\textrm{eff}\left(T\right)\sum_{i=1}^{n} a_i \int_{\tilde{d}} m\left(H_0,\theta_H, \varphi_H, \varphi_{ci}, \theta_{ci}, \psi_{ci},\tilde{d}\right) g_\textrm{TEM}(\tilde{d}) d\tilde{d}.
\label{eq:Mtot2}
\end{eqnarray}

\noindent where $M_\textrm{eff}(T) = M_s(T)\frac{V_n}{V_e}$.

\end{widetext}

\section{Analysis of results and discussion}

Using the model described above, we have calculated the effect of the field strength on the magnetic response of our GaP:MnP epilayers at $T$ = 180 K. Known parameters of bulk MnP at $T$ = 180 K, $M_s$ = 385 kA/m, $K_1$ = 871 kJ/m$^3$ and $K_2$ = 266 kJ/m$^3$ were used in the calculations.~\cite{huber2} The fitting curves as obtained from size distributions were used for the appropriate sample (see Fig.~\ref{fig:TEM}). The $a_i$ coefficients and $\theta_c$, $\varphi_c$ and $\psi_c$ angles were determined from ferromagnetic resonance measurements.~\footnote{The coefficients for sample GMP650 can be found in the eighth column of Table I of Ref.~\onlinecite{lacroix2013}.} The sole parameters that were fitted in the calculations are $V_n$, the volume occupied by the MnP nanoclusters, and $K_s$, a surface anisotropy constant. 

In Fig.~\ref{fig:hystnoHs}, we compare the experimental curve (circles) measured on sample GMP600 at $T$ = 180 K with the calculated curve (including thermal effects) when no surface anisotropy is included ($K_s = 0$, dashed line) and when $K_s = 0.54\times10^{-3}$ J/m$^2$ (solid line). When a surface anisotropy is included, excellent agreement between the experiments and calculations is found. For comparison, the calculated hysteresis curve when no surface anisotropy, nor thermal fluctuations are included is also presented (dotted line). The model was used to reproduce the hysteresis curves of all three samples for different directions of the magnetic field relative to the GaP substrate (solid lines in Fig.~\ref{fig:hystGMP}). We note that the values obtained for $V_n$ are in good agreement with values obtained from analyses of TEM micrographs (4~-~7~\%).~\cite{lambert2012} The values of $K_s$ obtained (see Table~\ref{tab:param}) are in the range of what is generally reported in the case of surface anisotropy.~\cite{coey2009}

The fact that the energy barrier required to fit the experiment is much lower than that calculated from the original Stoner-Wolhfarth model including thermal fluctuations indicates that magnetization reversal is non uniform. Our results show that the mechanism that induces magnetization reversal is a surface mechanism. This is supported by the fact that the use of size distributions obtained from TEM in combination with a surface contribution allow us to reproduce the slope of the hysteresis curve during magnetization reversal for all samples and different magnetic directions. As demonstrated earlier by micromagnetic calculations,~\cite{zhang1996,kachkachi2002} a reduction in coercivity can be observed in the presence of a Néel-type surface anisotropy. Our model also explains why the coercive field is the lowest in case of sample GMP600 (smallest average diameter) and the highest in case of sample GMP700 (largest average diameter). 

\begin{table} 
\caption{\label{tab:param} Surface anisotropy constants and MnP volume fraction values used to model the hysteresis curves of GaP:MnP epilayers at $T$ = 180 K. The  superscript number appearing with the symbol $K_s$ indicates the GaP crystallographic direction in which the magnetic field was applied. The error on the $K_s$ is estimated to be $\pm$ 0.02$\times 10^{-3}$ J/m$^2$.}
\begin{ruledtabular} 
\begin{tabular} {cccccc}
Sample & $K_s^{110}$& $K_s^{100}$ & $K_s^{1\bar{1}0}$ & $K_s^{001}$ & $V_n/V_e$\\
name &  (mJ/m$^2$)  &  (mJ/m$^2$)  &  (mJ/m$^2$)  &  (mJ/m$^2$)  & \\
\hline \\
GMP600  & 0.64 & 0.58 & 0.58 & 0.54 & 0.075\\
GMP650 & 0.72 & 0.58 & 0.72 & 0.58 & 0.067\\
GMP700 & 0.72 & 0.58 & 0.68 & 0.56 & 0.040\\
\end{tabular}
\end{ruledtabular}
\end{table}

\begin{figure} 
\begin{center}
\includegraphics[width=0.5\textwidth]{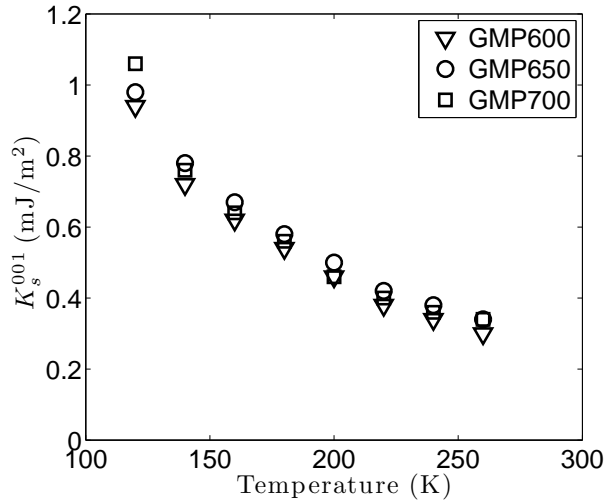}
\caption{\label{fig:KsvsT} Temperature dependence of $K_s^{001}$ for all three samples obtained from the fit of the experimental hysteresis curves. }
\end{center}
\end{figure}

From Table~\ref{tab:param}, we observe that the value of $K_s$ is the same (within the error) for all samples when the magnetic field is applied in the GaP $[001]$ and $[100]$ directions, while it varies depending up on the samples when the magnetic field is applied in the GaP $[110]$ and $[1\bar{1}0]$ directions. We explain this result as due to the symmetry of the GaP substrate and the differences in the volume fraction of MnP nanoclusters for each orientation.~\cite{lacroix2013} When the field is applied in the GaP $[001]$ and $[100]$ directions, because of symmetry considerations, several cluster orientations present the same hysteresis curve, giving a total of four distinct hysteresis curves, two of which have nonzero coercivity. Since almost all clusters (more than $\approx$ 90\%) showing nonzero coercivity present the same hysteresis curve, we are effectively probing MnP nanoclusters with their magnetic axis all aligned in the same direction. When the field is applied in the GaP $[110]$ and $[1\bar{1}0]$ directions, six distinct hysteresis curves are present, three of which have nonzero coercivity. We then have several populations of MnP nanoclusters with different orientations and volume fraction that contributes to the magnetization reversal, which give rise to different $K_s$ values.

Therefore, our results suggest that the surface anisotropy hardly changes as a function of growth temperature, \textit{i.e.} the structure of MnP, and most probably the MnP/GaP interface is not affected by the growth temperature. However, because the size distribution and the quantity of MnP nanoclusters along each crystallographic orientation are strongly affected by the growth temperature, we obtain very different hysteresis curves for samples grown at different temperatures.

Finally, we have measured the hysteresis curves of all three samples at various temperatures in the range from 120 to 260 K with the magnetic field applied in the GaP $[001]$ direction. Hysteresis curves could be well fitted with the present model and the values of $M_s$, $K_1$ and $K_2$ from bulk MnP for all temperatures. The temperature dependence of $K_s^{001}$, which is the $K_s$ value obtained when $H_0$ is applied in the GaP [001] direction, is presented in Fig.~\ref{fig:KsvsT}. We observe that $K_s^{001}$ increases by a factor of 3 as the temperature is lowered from 260 to 120 K. We note also that $K_s^{001}$ values obtained from the three samples are very close for all temperatures except at 120 K, where the surface contribution increases strongly. 

\section{Summary}

We have determined the quasi-static magnetic response of ensembles of non-interacting ferromagnetic nanoparticles embedded in a non magnetic medium. The Stoner-Wohlfarth model was extended to incorporate a orthorhombic anisotropy as well as a surface anisotropy contribution to the energy barrier. The model was successfully implemented to account for the measured hysteresis curves in the temperature range from 120 to 260 K, with essentially one adjustable parameter at each temperature, namely the surface anisotropy constant. Our results show that the slope of the magnetization curve during magnetization reversal is determined by the size distribution, which is a direct consequence of the presence of this surface (or interface) contribution. Finally, we have shown that the energy barrier between the two states of opposite magnetization is strongly reduced as the temperature is lowered from 260 to 120 K due to these surface effects. 

The authors would like to acknowledge A. Yelon for providing critical comments on the manuscript. This work was supported by grants from NSERC (Canada) and FRQNT (Québec).

\end{document}